\documentclass[conference,twocolumn, 10pt]{IEEEtran}
\usepackage{epsfig, latexsym, amssymb, amsmath}
\usepackage{graphics}
\usepackage{algorithm}
\usepackage{algorithmic}
\IEEEoverridecommandlockouts

\newtheorem{prop}{Proposition}[section]

\newtheorem{cor}{Corollary}

\newtheorem{lm}{Lemma}

\newtheorem{thm}{Theorem}

\newcommand{\bthm}{\begin{thm}}
\newcommand{\ethm}{\end{thm}}

\newcommand{\bcor}{\begin{cor}}
\newcommand{\ecor}{\end{cor}}
\newcommand{\bprop}{\begin{prop}}
\newcommand{\eprop}{\end{prop}}
\newcommand{\blm}{\begin{lm}}
\newcommand{\elm}{\end{lm}}
\newcommand{\beq}{\begin{equation}}
\newcommand{\eeq}{\end{equation}}
\newcommand{\ber}{\begin{eqnarray}}
\newcommand{\eer}{\end{eqnarray}}

\newenvironment{proof1}{\begin{trivlist}\item[]{\bf Proof:\hspace{2mm}}}{\hfill$\blackbox$\end{trivlist}}
\newcommand{\sr}[1]{(\ref{#1})}

\newcommand{\argmax}{\mathop{\mbox{\rm arg\,max}}}

\newcommand{\blackbox}{\vrule height7pt width5pt depth1pt}

\newcommand{\bit}{\begin{itemize}}
\newcommand{\eit}{\end{itemize}}
\newcommand{\ben}{\begin{enumerate}}
\newcommand{\een}{\end{enumerate}}
\newcommand{\bdesc}{\begin{description}}
\newcommand{\edesc}{\end{description}}
\newcommand{\beqarrn}{\begin{eqnarray*}}
\newcommand{\eeqarrn}{\end{eqnarray*}}
\newenvironment{proofof}[1]{\begin{trivlist}\item[]{\bf Proof of #1:\hspace{2mm}
}}{\hfill\blackbox\end{trivlist}}
\newcommand{\bproofof}{\begin{proofof}}
\newcommand{\eproofof}{\end{proofof}}
\newenvironment{rem}{\begin{trivlist}\item[]{\bf
Remark:}\hspace{4mm}}{\end{trivlist}}
\newcommand{\brem}{\begin{rem}}
\newcommand{\erem}{\end{rem}}
\newenvironment{rems}{\begin{trivlist}\item[]{\bf
Remarks}\begin{itemize}}{\end{itemize}\end{trivlist}}
\newcommand{\brems}{\begin{rems}}
\newcommand{\erems}{\end{rems}}
\newtheorem{fact}{Fact}
\newcommand{\bfact}{\begin{fact}}
\newcommand{\efact}{\end{fact}}
\newtheorem{examp}{Example}
\newcommand{\bexamp}{\begin{examp}\rm}
\newcommand{\eexamp}{\end{examp}}
\newtheorem{defn}{Definition}
\newcommand{\bdefn}{\begin{defn}\rm}
\newcommand{\edefn}{\end{defn}}

\newtheorem{prob}{Problem}
\newcommand{\bprob}{\begin{prob}}
\newcommand{\eprob}{\end{prob}}

\newcommand{\bvtm}{\begin{verbatim}}
\newcommand{\bfig}{\begin{figure}}
\newcommand{\efig}{\end{figure}}
\newcommand{\bcen}{\begin{center}}
\newcommand{\ecen}{\end{center}}

\long\def\comment#1{}

\def \n2{{N_0 \over 2}}

\def \h5{\hspace{0.5in}}

\begin{document}

\title{\LARGE Distributed Opportunistic Scheduling For Ad-Hoc
Communications Under Noisy Channel Estimation}
\author{Dong Zheng, Man-On Pun, Weiyan Ge, Junshan Zhang and H. Vincent Poor
\thanks{Dong Zheng is with NextWave Wireless Inc., San Diego, CA 92130 (e-mail: dzheng@nextwave.com).}
\thanks{Man-On Pun and H. Vincent Poor are with the Department of Electrical Engineering, Princeton University, Princeton, NJ 08544 (e-mail: mopun@princeton.edu; poor@princeton.edu).}
\thanks{Weiyan Ge and Junshan Zhang are with the Department of Electrical Engineering, Arizona State University, Tempe, AZ 85287 (e-mail: Weiyan.Ge@asu.edu; Junshan.Zhang@asu.edu).}
\thanks{This research was supported in part by the Croucher Foundation under a post-doctoral fellowship, in part by the U. S. National Science Foundation under Grants ANI-02-38550, ANI-03-38807, CNS-06-25637, and CNS-07-21820 and in part by Office of Naval Research through Grant
N00014-05-1-0636.}}
\maketitle

\begin{abstract}
Distributed opportunistic scheduling is studied for wireless ad-hoc
networks, where many links contend for one channel using random
access. In such networks, distributed opportunistic scheduling (DOS)
involves a process of joint channel probing and distributed
scheduling. It has been shown that under perfect channel estimation,
the optimal DOS for maximizing the network throughput is a pure
threshold policy. In this paper, this formalism is generalized to
explore DOS under noisy channel estimation, where the
transmission rate needs to be backed off from the estimated rate to
reduce the outage. It is shown that the optimal scheduling policy
remains to be threshold-based, and that the rate threshold turns out
to be a function of the variance of the estimation error and be a
functional of the backoff rate function. Since the optimal backoff
rate is intractable, a suboptimal linear backoff scheme that backs off
the estimated signal-to-noise ratio (SNR) and hence the rate is
proposed. The corresponding optimal backoff ratio and rate
threshold can be obtained via an iterative algorithm. Finally,
simulation results are provided to illustrate the tradeoff caused by
increasing training time to improve channel estimation at the cost
of probing efficiency.
\end{abstract}

\section{Introduction}
Channel-aware scheduling for achieving rich diversities inherent in
wireless communications has recently emerged as a promising
technique for improving spectral efficiency in wireless networks.
Most existing studies along this line require centralized scheduling
(see \cite{qin03} and the
references therein), and little work has been done on developing
distributed algorithms to harvest diversity gains for ad hoc
communications. Unlike centralized cases, in ad hoc
communication networks, each link has no knowledge of other links'
channel conditions, making it very challenging to carry out
channel-aware distributed scheduling.

Some initial steps have been taken by several of the authors to develop
distributed opportunistic scheduling (DOS) to reap multiuser
diversity and time diversity in wireless ad hoc networks \cite{
zgz:mobihoc}. In DOS, a successful link proceeds to data transmission only if
the observed channel condition is ``good''; otherwise, it may skip the
transmission, and let all the links re-contend for the channel.
Intuitively speaking, because different links in different time
slots experience different channel conditions, it is likely that
after further probing, the channel can be taken by a link with a
better channel condition, resulting in possible higher throughput.
In this way, the multiuser diversity across links and the time
diversity across slots can be exploited in a joint manner.

Despite the insightful analytical results reported in~\cite{zgz:mobihoc},
its key assumption is that perfect channel state information (CSI) is
known at the receiver/transmitter. However, in practical
scenarios, channel conditions are often estimated using noisy
observations. Therefore, it is of great interest to study DOS under
noisy channel estimation. In centralized scheduling schemes, it has
been shown that the estimated signal-to-noise ratio (SNR) is always
larger than the ``actual SNR''~\cite{vakili06}. Thus, if the data
were transmitted using the estimated rate, there would always be an
outage. To reduce the outage probability, the transmission rate has
to back off from the estimated rate. Therefore, unlike the perfect
estimation case, the optimal scheduling policy hinges on the backoff
rate.

In this work, we generalize~\cite{zgz:mobihoc} to the scenario with
imperfect channel estimation and show that the optimal scheduling
policy for the noisy channel estimation case remains a threshold
structure. However, the threshold turns out to be a function of the
variance of the channel estimation error, and furthermore, it is a
functional of the backoff rate function. Since the optimal backoff
rate function is difficult to obtain, we propose a suboptimal linear
backoff scheme. We show that the corresponding optimal
backoff ratio and rate threshold can be obtained via an iterative
numerical algorithm. Simulation results are provided to show that
DOS achieves significant throughput gain in the presence of noisy
channel estimation, especially in the low SNR region.

\section{System Model and Background}\label{sec:rw}
We consider a single-hop ad hoc
network with $M$ links, where link $m$
contends for the channel with probability $p_m$, $m =1, \ldots, M$.
A collision model is assumed for random access, where channel
contention of a link is said to be successful if no other links
transmit at the same time. Accordingly, the overall successful
contention probability, $p_s$, is then given by $ \sum_{m=1}^M (p_m
\prod_{i \neq m} (1-p_i)) $. It is clear
that the number of slots (denoted as $K$)  for a successful channel
contention is a Geometric random variable (r.v.), i.e., $K \sim
Geometric(p_s)$. Let $\tau$ denote the duration of mini-slot for
channel contention, and $T$ the data transmission time. It follows
that the random duration corresponding to one round of successful
channel contention is $K \tau$, with expectation $\tau/p_s$.

Let $s(n)$ denote the successful link at the $n$-th successful
channel contention. The corresponding received signal is given by:
\begin{equation} \label{eq:channel}
Y_{s(n)}(n) = \sqrt{\rho}h_{s(n)}(n)X_{s(n)}(n) + \mu_{s(n)}(n),
\end{equation}
where $\rho$ is the normalized receiver SNR, $h_{s(n)}(n)$ is the
channel coefficient for link $s(n)$, $X_{s(n)}(n)$ is the
transmitted signal with $E\left\{||X_{s(n)}(n)||^2\right\}=1$ and $\mu_{s(n)}(n)$
is additive white noise with i.i.d. $\mathcal{CN}(0,1)$.

To simplify the exposition, we consider a homogeneous network in which
all links have the same channel statistics, and are subject to
Rayleigh fading, i.e., $h_{s(n)}(n)$ follows a  complex Gaussian
distribution $\mathcal{CN}(0,1)$. In what follows, we drop the
subscripts to simplify the notation and use $h_n$ to stand for
$h_{s(n)}(n)$ where it is clear from the context. Similarly, we use
$Y_n$, $X_n$ and $\mu_n$ to denote $Y_{s(n)}(n), X_{s(n)}(n)$ and
$\mu_{s(n)}(n)$.

We consider the continuous rate case, assuming that the
instantaneous rate is given by the Shannon channel capacity, i.e.,
\begin{equation*}
  R_n = \log(1+\rho |h_n|^2) \ \textrm{nats/s/Hz},
  \end{equation*}
provided that the channel can be perfectly estimated.

In~\cite{zgz:mobihoc}, we have studied DOS with perfect CSI.
Specifically, we have shown that the problem
can be cast as a {\em maximal rate of return} problem in optimal
stopping theory~\cite{ferguson:os}, where the rate of return is the
average network throughput, $x$, and is determined by the stopping
time $N$:
\begin{equation}
\begin{small}
  x =  \frac{E[R_{N}T]}{E[T_N]}, \label{eq:ob}
\end{small}
\end{equation}
where $T_n \triangleq \sum_{j=1}^{n}K_{j}\tau+T$ is the total system
time including the contention time and the data transmission time.
Note that $N$ is a stopping time if $\{N=n\}$ is
$\mathcal{F}_n$-measurable, where $\mathcal{F}_n$ is the
$\sigma$-field generated by $\{(\rho|h_j|^2, K_j),
j=1,2,\ldots,n\}$.

We show that the optimal DOS maximizing the throughput is given by
the optimal stopping rule, $N^*$, that solves the maximal rate of
return problem in \sr{eq:ob}, i.e.,
\begin{small}
\begin{equation}
N^* \triangleq \argmax\limits_{N\in Q} \frac{E[R_N T]}{E[T_N]},
~~x^* \triangleq \sup_{N\in Q} \frac{E[R_NT]}{E[T_N]},
\label{main-problem}
\end{equation}
\end{small}
where
\begin{small}
\begin{equation} \label{eq:q}
Q \triangleq \{N: N\ge 1, E[T_N]<\infty\}.
\end{equation}
\end{small}

It has been shown in \cite{zgz:mobihoc} that
the optimal stopping rule $N^*$ for DOS exists, and is given by
\begin{equation} \label{eq:stop}
  N^* = \min\{n \ge 1: R_{n} \ge x^*\}.
  \end{equation}
Furthermore, the maximal throughput $x^*$ is an optimal threshold, and is
the unique solution to
\begin{equation}
  E(R-x)^+ = \frac{x \tau}{p_sT}, \label{eq:rate}
  \end{equation}
where $R$ is a r.v. having the same distribution as
$R_{n}$.

For example, if $R_n= \log(1+\rho |h_n|^2)$ and $h_n$ has a complex
Gaussian distribution $\mathcal{CN}(0,1)$, it can be shown that
\begin{eqnarray}
  x^* =
{\exp\left(\frac{1}{\rho}\right)E_1\left(\frac{\exp(x^*)}{\rho}\right)}{\frac{p_s}{\delta}},
\label{eq:crate}
  \end{eqnarray}
where $\delta=\tau/T$, and $E_1(x)$ is the \emph{exponential
integral function} defined as $E_1(x) \triangleq \int_x^{\infty} \frac{\exp(-t)}{t}dt$.

\section{DOS Under Noisy Channel Estimation} \label{sec:ndos}
Needless to say, in practical systems, $h_n$ has to be estimated
using training signals (e.g. embedded in the RTS packets). Let
$\hat{h}_n$ denote the estimation of the channel coefficient, and
$\tilde{h}_n$ the estimation error. It follows that
\begin{equation} \label{eq:hn}
h_n = \hat{h}_n + \tilde{h}_n,
\end{equation}
where $\hat{h}_n$ and $\tilde{h}_n$ are zero-mean complex Gaussian
random variables. Suppose that the channel is estimated using a
minimum mean square error (MMSE)-based estimator. It follows, by the orthogonality principle, that
\begin{equation} \label{eq:orth}
E[|h_n|^2]=E[|\hat{h}_n|^2] + E[|\tilde{h}_n|^2].
\end{equation}
Let $\beta$ denote the variance of the estimation error. From
\sr{eq:orth}, we have that
\begin{equation}
E[|\tilde{h}_n|^2]  =  \beta \quad\mbox{and}\quad E[|\hat{h}_n|^2]  =  1- \beta.
\end{equation}

Treating the estimation error as noise, the actual SNR at the
receiver can be computed by~\cite{vakili06}
\begin{equation} \label{eq:acsnr}
\lambda_n = \frac{\rho|\hat{h}_n|^2}{1+\rho|\tilde{h}_n|^2}.
\end{equation}
We note that the numerator of \sr{eq:acsnr}, $\rho|\hat{h}_n|^2$, is
the estimated SNR. Therefore, in contrast to the perfect CSI case
where the sequence $\{\rho|h_n|^2, n=1,2,\ldots \}$ is used for
distributed scheduling, in the noisy estimation case,
$\{\rho|\hat{h}_n|^2, n=1,2,\ldots \}$ serves as the basis for
distributed scheduling.

Following~\cite{vakili06}, $|\hat{h}_n|^2$ and $|\tilde{h}_n|^2$ can
be normalized as
\begin{equation}
\hat{\lambda}_n = \frac{|\hat{h}_n|^2}{1-\beta} \quad\mbox{and}\quad z_n =
\frac{|\tilde{h}_n|^2}{\beta}.
\end{equation}
Note that both $\hat{\lambda}_n$ and $z_n$ have the exponential
distribution with unit variance. Furthermore, $\lambda_n$ in
\sr{eq:acsnr} can be rewritten as
\begin{equation} \label{eq:nacsnr}
\lambda_n = \frac{\rho_{eff}\hat{\lambda}_n}{1+\alpha\rho_{eff}z_n},
\end{equation}
where $\rho_{eff}\triangleq (1-\beta)\rho$ and $\alpha \triangleq
\frac{\beta}{1-\beta}$ denote the ``effective channel SNR'' and
``normalized error variance'', respectively. It can be shown the
distribution of $\lambda_n$ given $\hat{\lambda}_n$ is given
by~\cite{vakili06}
\begin{small}
\begin{equation} \label{eq:conlam}
f\left(\lambda_n | \hat{\lambda}_n\right) =
\frac{\hat{\lambda}_n}{\alpha \lambda^2_n}
\exp\left\{-\frac{1}{\alpha}\left(\frac{\hat{\lambda}_n}{\lambda_n}-\frac{1}{\rho_{eff}}\right)\right\}
\mathbf{I}\left(\frac{\hat{\lambda}_n}{\lambda_n}-\frac{1}{\rho_{eff}}\right),
\end{equation}
\end{small}
where $\mathbf{I}(\cdot)$ is the indicator function.

\subsection{Optimal Stopping Rule under Noisy Channel Estimation}

It is clear that the actual SNR $\lambda_n$ is no greater than the
estimated SNR $\rho_{eff}\hat{\lambda}_n$. As a result, if the
packet is transmitted at the estimated rate
$\log(1+\rho_{eff}\hat{\lambda}_n)$, there would always be a channel
outage. Therefore, the transmission rate has to back off from the
estimate rate. Equivalently, we can back off the estimated SNR
$\rho_{eff}\hat{\lambda}$ to a ``nominated'' SNR
$\lambda_c(\hat{\lambda})$. Accordingly, the instantaneous rate,
$R_n$, is given by
\begin{equation} \label{eq:inrate}
R_n =
\log\left(1+\lambda_c(\hat{\lambda}_n)\right)\mathbf{I}\left(\lambda_c(\hat{\lambda}_n)
\le \lambda_n\right).
\end{equation}

Along the same line as in the perfect CSI case, for each given
back-off rate function $\lambda_c(\cdot)$, maximizing the average
throughput reduces to solving the maximal rate of return problem
in \sr{eq:ob}.

Observe that there are at least two major differences between the
perfect estimation case and the noisy channel estimation case.
First, the stopping rule $N$ is now defined over the $\sigma$-field
$\mathcal{F}'_n$ (instead of $\mathcal{F}_n$), generated by
$\{(\rho|\hat{h}_j|^2, K_j), j=1,2,\ldots,n\}$. Second, the
instantaneous rate, $R_n$, defined in \sr{eq:inrate}, is now a
random variable, and is not perfectly known at time $n$. However, it
can be shown that the structure of the optimal scheduling strategy
remains the same, except that the random ``reward'' $R_n$ is
replaced with its conditional expectation, $\bar{R}_n \triangleq
E\left[R_n| \mathcal{F}'_n\right]$~\cite[Page
1.3]{ferguson:os}~\cite{assaf98}. More specifically, define
\begin{equation}
Q' \triangleq \{N \ge 1: \{N=n\}\in \mathcal{F}'_n, E[T_N]<\infty\}.
\end{equation}
Analogously, define
\begin{equation}
Q'' \triangleq \{N\ge 1: \{N=n\}\in \mathcal{F}''_n,
E[T_N]<\infty\},
\end{equation}
where $\mathcal{F}''_n$ is the $\sigma$-field generated by
$\{(\bar{R}_j, K_j), j=1,2,\ldots,n\}$. We have the following
proposition. \bprop \label{prop:eq}
\begin{equation}
\sup_{N\in Q'}\frac{E[R_NT]}{E[T_N]} = \sup_{N \in Q''}
\frac{E[\bar{R}_NT]}{E[T_N]}.
\end{equation}
\eprop The proof follows from Propositions 2.3, 2.4, 2.5
in~\cite{assaf98} (with $X_n, Z_n, W_n$ in lieu of $R_n,
\hat{\lambda}_n, \bar{R}_n$), and the proof in~\cite{zgz:mobihoc}.
Due to space limitation, we omit the details here.

As a result, Proposition~\ref{prop:eq} indicates that the optimal
scheduling can be based solely on $\bar{R}_n$, given by
\begin{small}
\begin{eqnarray}
\bar{R}_n &=&  E\left[R_n| \mathcal{F}'_n\right]\nonumber\\
&=&\log\left(1+\lambda_c(\hat{\lambda}_n)\right)\left[1-\exp\left\{-\frac{1}{\alpha}\left(\frac{\hat{\lambda_n}}{\lambda_c(\hat{\lambda}_n)}
- \frac{1}{\rho_{eff}}\right)\right\}\right], \nonumber
\end{eqnarray}
\end{small}
where we have used the fact that $P\left(\lambda_c(\hat{\lambda}_n)
\le \lambda_n|\mathcal{F}'_n\right) =
P\left(\lambda_c(\hat{\lambda}_n) \le
\lambda_n|\hat{\lambda}_n\right)$ due to the independence of channel
estimations.

Based on the above discussion, we conclude that the optimal
scheduling policy under noisy channel estimation is a pure threshold
policy:
\begin{equation} \label{eq:bstop}
  N^* = \min\{n \ge 1: \bar{R}_{n} \ge x^*\},
  \end{equation}
where the optimal threshold $x^*$ can be computed from \sr{eq:rate},
and hence, it is the unique solution to the following fixed point
equation:
\begin{equation}
x  = \Phi(x, \lambda_c), \label{eq:phirhor}
\end{equation}
where
\begin{small}
\begin{equation}
   \Phi(x, \lambda_c)  \triangleq  \frac{\int_{\hat{\lambda}'}^{\infty} e^{-\hat{\lambda}}\log\left(1+\lambda_c\right)\left[1-\exp\left\{-\frac{1}{\alpha}\left(\frac{\hat{\lambda}}{\lambda_c}
- \frac{1}{\rho_{eff}}\right)\right\}\right] d\hat{\lambda}}{
  \frac{\delta}{p_s}+e^{-\hat{\lambda}'}}, \label{eq:p1}
\end{equation}
\end{small}
and $\hat{\lambda}'$ can be obtained from
\begin{small}
\begin{equation} \label{eq:intrange}
\log\left(1+\lambda_c(\hat{\lambda}')\right)\left[1-\exp\left\{-\frac{1}{\alpha}\left(\frac{\hat{\lambda}'}{\lambda_c(\hat{\lambda}')}
- \frac{1}{\rho_{eff}}\right)\right\}\right] = x.
\end{equation}
\end{small}

\subsection{Optimal Backoff Rate Function}
It is clear from \sr{eq:phirhor} that for a given backoff rate
function $\lambda_c(\cdot)$, there is a corresponding optimal
throughput $x^*$. Therefore, $x^*$ is a functional of
$\lambda_c(\cdot)$, denoted as $x^*(\lambda_c)$. We are interested
in finding the function $\lambda^*_c(\cdot)$ that maximizes
$x^*(\lambda_c)$, i.e.,
\begin{equation} \label{eq:calv}
\lambda^*_c = \argmax\limits_{\lambda_c \in A} x^*(\lambda_c),
\end{equation}
where $A$ is the set of the admissible functions (for example, $A$
can be $\{\lambda_c(\hat{\lambda}): \lambda_c(\hat{\lambda}) \ge 0,
\forall \ \hat{\lambda} \ge 0 \}$).

Based on the theory of calculus of variations~\cite{gelfand_fomin},
problem \sr{eq:calv} is a \emph{variational problem}, and the
functions $\lambda^*_c(\cdot)$ are called \emph{extremals}. However,
unlike the canonical calculus of variations problems, in
this problem, the functional $x^*$ is not explicitly defined on
$\lambda_c$. Instead, they are connected through a fixed point
equation. Furthermore, the integral range in \sr{eq:p1} is not
fixed, but is a function of $\lambda_c$ (cf. \sr{eq:intrange}). As a
result, it is intractable to characterize $\lambda^*_c$.

\subsection{A Suboptimal Backoff Rate Function}
In what follows, we propose a suboptimal backoff rate function,
which backs off the estimated SNR by a multiplicative ratio
$\sigma$, i.e., we set
\begin{equation} \label{eq:subop}
\lambda_c(\hat{\lambda}) = \sigma \rho_{eff}\hat{\lambda},
\end{equation}
and $ 0\le \sigma \le 1$.

It follows from \sr{eq:subop}, \sr{eq:p1} and \sr{eq:intrange} that
the optimal throughput $x^*$ is the solution to
\begin{small}
\begin{eqnarray}
&& \mbox{} \hspace{-0.98cm}  x = \Phi(x, \sigma) \label{eq:phisigma} \nonumber \\
  && \mbox{} \hspace{-0.8cm}=  \left[1-\exp\left\{-\frac{1}{\alpha\rho_{eff}}\left(\frac{1}{\sigma}
- 1\right)\right\}\right] \nonumber
\\
&& \mbox{} \hspace{-0.8cm} \times
\frac{\log(1+\sigma\rho_{eff}\hat{\lambda}')e^{-\hat{\lambda}'}+\exp\left(\frac{1}{\sigma
\rho_{eff}}\right)E_1\left(\hat{\lambda}' + \frac{1}{\sigma
\rho_{eff}}\right)}{
  \frac{\delta}{p_s}+e^{-\hat{\lambda}'}}, \label{eq:p3}
\end{eqnarray}
\end{small}
where
\begin{small}
\begin{equation} \label{eq:lambdaprime}
\hat\lambda' =
\frac{\exp\left(\frac{x}{1-\exp\left\{-\frac{1}{\alpha\rho_{eff}}\left(\frac{1}{\sigma}-1\right)\right\}}\right)
-1}{\sigma \rho_{eff}}.
\end{equation}
\end{small}

It is not difficult to show that $x^*$ is a continuous and
differentiable function of $\sigma$, and hence, there exists an
optimal backoff ratio $\sigma^*$ such that
\begin{equation} \label{eq:osigma}
\sigma^* = \argmax_{\sigma}x^*(\sigma).
\end{equation}
It can also be shown that $\sigma^*$ cannot be $0$ or $1$ (since the
corresponding throughput is zero). Therefore, the optimal ratio
$\sigma^*$ must satisfy the first order condition
$\frac{dx^*(\sigma^*)}{d\sigma} = 0$.

\subsection{An Iterative Algorithm for Computing $\sigma^*$ and
$x^*(\sigma^*)$} Due to the complicated structure of the fixed point
equation \sr{eq:p3}, it is not feasible to characterize $\sigma^*$
using the first order condition. In what follows, we devise an
iterative algorithm instead using fractional optimization
techniques~\cite{bertsekas:np}.

Specifically, we define the following functions:
\begin{small}
\begin{eqnarray*}
&&\mbox{} \hspace{-0.8cm} U(\sigma, x)\triangleq
\left[1-\exp\left\{-\frac{1}{\alpha\rho_{eff}}\left(\frac{1}{\sigma}
- 1\right)\right\}\right] \times \nonumber
\\
&& \mbox{} \hspace{-0.8cm}
\left\{\log(1+\sigma\rho_{eff}\hat{\lambda}')\exp(-\hat{\lambda}')+
 \exp\left(\frac{1}{\sigma
\rho_{eff}}\right)E_1\left(\hat{\lambda}' + \frac{1}{\sigma
\rho_{eff}}\right)\right\},
\end{eqnarray*}
\end{small}
and
$
 V(\sigma, x) \triangleq
\frac{\delta}{p_s}+\exp(-\hat{\lambda}'),
$ where $\hat{\lambda}'$ is defined in \sr{eq:lambdaprime}.

The iterative algorithm is outlined in Algorithm~\ref{al:itr}, and its convergence is established in \cite{dong2007}.

\begin{algorithm}
\caption{Iterative Algorithm for Computing $\left\{\sigma^*,x^*(\sigma^*)\right\}$} \label{al:itr}
\begin{algorithmic}
\STATES
\STATE $x_k$, $\sigma_k$
%\INIT \STATE $x_0$
\PROCEDURE
\WHILE{$|x_k-x_{k-1}|>\epsilon$}
\STATE $\sigma_{k-1} = \argmax\limits_{0\le \sigma \le 1}
\left\{U(\sigma,x_{k-1})-x_{k-1}V(\sigma,x_{k-1})\right\}$
\STATE $x_k =
\frac{U(\sigma_{k-1},x_{k-1})}{V(\sigma_{k-1},x_{k-1})}$ \ENDWHILE
\end{algorithmic}
\end{algorithm}

\section{Numerical Results} \label{sec:app}
In this section, we provide numerical examples to illustrate the
above results. Unless otherwise specified, we assume that $\tau$,
$T$, $p$, and $M$ are chosen such that $\delta = 0.1$ and $p_s=\exp(-1)$.

\begin{figure}[htp]
\begin{center}
\includegraphics[width=0.4\textwidth]{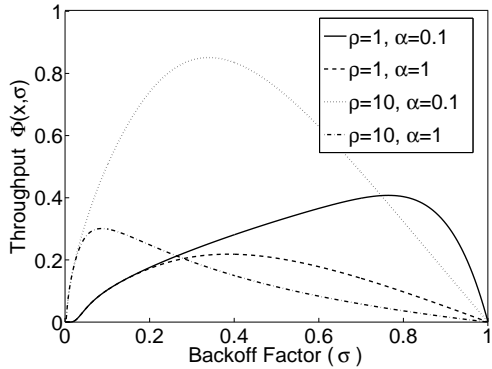}
\end{center}
\caption{$\Phi(\sigma)$ vs. $\sigma$, $x=0.1$.} \label{fig:th_sigma}
\end{figure}

Figure \ref{fig:th_sigma} depicts $\Phi(x,\sigma)$ as a function of
the backoff ratio $\sigma$. It can be seen that the average
throughput is zero at both $\sigma=0$ and $\sigma=1$, and is
maximized somewhere in between.

\begin{figure}[htp]
\begin{center}
\includegraphics[width=0.4\textwidth]{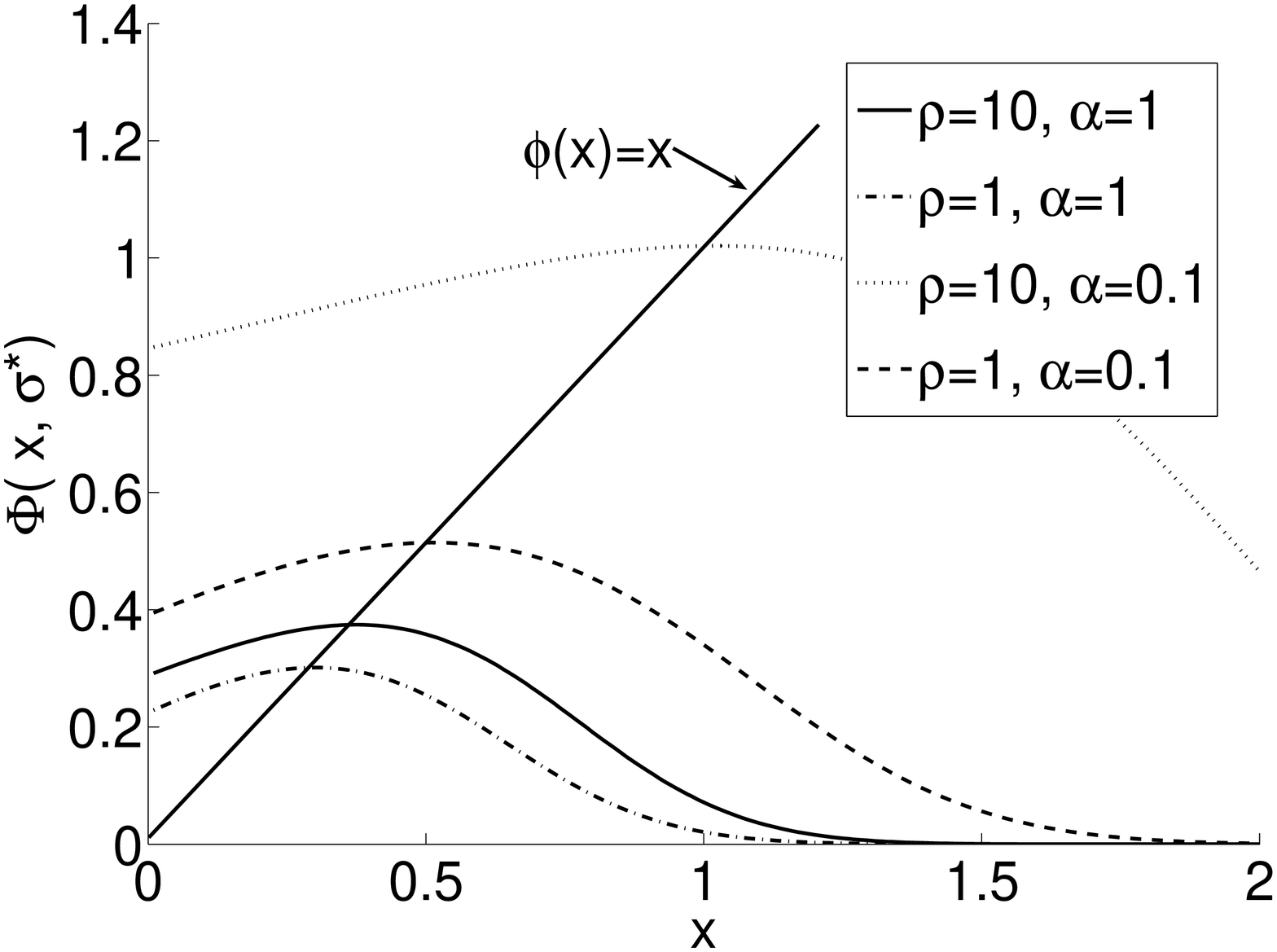}
\end{center}
\caption{$\Phi(x,\sigma^*)$ vs. $x$} \label{fig:th_x}
\end{figure}

Figure~\ref{fig:th_x} depicts $\Phi(x,\sigma^*)$ as a function of
$x$. Note that the optimal throughput $x^*$ is the solution to the
fixed point equation $x=\Phi(x,\sigma^*)$. It can be observed that
$x^*$ is an increasing function of $\rho$ for a given $\alpha$, and
is a decreasing function of $\alpha$ for a fixed $\rho$. It can also
be seen that the estimation accuracy plays an important role in the
throughput performance: when $\alpha$ decreases from $1$ to $0.1$,
the performance improves over $150\%$ for $\rho=10$.

In Table~\ref{t:con}, we examine the convergence of the iterative
algorithm I with $\alpha=1$. As expected, $x(n)$ approaches to $x^*$
usually within a few iterations.

\begin{table}[htp]
\caption{Convergence behavior of the iterative algorithm,
$\alpha=1$.}
\begin{center}
\begin{tabular}{| c | c | c | c | c | c |c|}
\hline
$\rho$ & $x_0$ &  $x_1$ &  $x_2$  &   $x_3$   & $x^*$ & $\sigma^*$\\
\hline
0.5& 0.5 & 0.177 & 0.246 & 0.254 & 0.254 & 0.407 \\
\hline
1& 0.5 & 0.254 & 0.299 & 0.301 & 0.301 & 0.285 \\
\hline
2 & 0.5 & 0.306 & 0.335 & 0.336  & 0.336 & 0.182 \\
\hline
5 & 0.5 & 0.344 & 0.363 & 0.364 & 0.364 & 0.090 \\
\hline
10 & 0.5 & 0.358 & 0.374 & 0.374 & 0.374 & 0.049 \\

\hline
\end{tabular}
\end{center}
\label{t:con}
\end{table}

Table~\ref{t:con_fixrho} compares the convergence behavior of the
iterative algorithm with different error variance $\alpha$ and $\rho=1$.
When the error variance is large, the
iterative algorithm needs more iterations to converge. Moreover, the
backoff ratio $\sigma$ would decrease as $\alpha$ increases. This
can be further observed in Fig. \ref{fig:backoff_error}. It
indicates that when the estimation error is large, the transmitter
would back off more to avoid channel outage.

\begin{table}[htp]
\caption{Convergence behavior of the iterative algorithm, $\rho=1$.}
\begin{center}
\begin{tabular}{| c | c | c | c | c | c |c|c|c|}
\hline
$\alpha$ & $x_0$ &  $x_1$ &  $x_2$  &   $x_3$  &  $x_4$ & $x_5$ & $x^*$ & $\sigma^*$\\
\hline
0  & 0.5&  0.604 & 0.610  & && & 0.610 & 1.00 \\
\hline
0.1 & 0.5&  0.514 & 0.514 & &&& 0.514 & 0.753 \\
\hline
1& 0.5 & 0.254 & 0.299 & 0.301 & && 0.301 & 0.285 \\
\hline
2 & 0.5 &   0.109 & 0.201 & 0.217 & 0.218 & & 0.218 & 0.155 \\
\hline
5 & 0.5 &  0.004 & 0.091 & 0.120 & 0.122 & 0.123 & 0.123 & 0.054 \\
\hline
\end{tabular}
\end{center}
\label{t:con_fixrho}
\end{table}

\begin{figure}[htp]
\begin{center}
\includegraphics[width=0.4\textwidth]{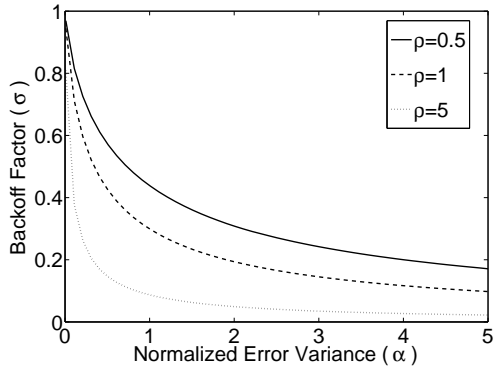}
\end{center}
\caption{Backoff factor $\sigma$ as a function of normalized error
variance $\alpha$} \label{fig:backoff_error}
\end{figure}

Table~\ref{t:gain_fixbeta} illustrates the throughput gain,
$g = \frac{x^*-x^L}{x^L}$, as a function of $\rho$, where $x^L = \Phi(0, \sigma^*)$ is
the average throughput obtained by the schemes without
 using optimal scheduling. It can be seen that the throughput gain
is more significant in the low SNR region, and is a decreasing
function of $\rho$.

\begin{table}[htp]
\caption{Throughput gain of DOS, $\alpha=1$.}
\begin{center}
\begin{tabular}{| c | c | c | c | c | c |c|}
\hline
$\rho$ & 0.5 &  1 &  2 &  5 &  10 & 100\\
\hline
$x^*$& 0.254 & 0.301 & 0.336 & 0.364 & 0.374 & 0.385\\
\hline
$x^L$ & 0.185 & 0.224 & 0.254 & 0.278 & 0.288 & 0.298\\
\hline
$g(\rho)$ & 37.3\% &  34.3\% & 32.3\% & 30.9\% & 29.8\% & 29.2\%\\
\hline
\end{tabular}
\end{center}
\label{t:gain_fixbeta}
\end{table}

In Table~\ref{t:gain_fixrho}, we illustrate the throughput gain as a
function of $\alpha$. Note that $\alpha = 1/(1-\beta)-1$ is an
increasing function of $\beta$. As expected, when the normalized
noise variance $\alpha$ increases, the optimal throughput $x^*$
decreases, as well as the $x^L$. However, it is interesting to
observe that the throughput gain increases instead. The rationale
behind is that the performance of the schemes that do not use optimal
scheduling ``suffers'' more than that of the distributed
opportunistic scheduling in the presence of noisy channel
estimation.

\begin{table}[htp]
\caption{Throughput gain of DOS, $\rho=0.5$.}
\begin{center}
\begin{tabular}{| c | c | c | c | c | c |c|}
\hline
$\alpha$ & 0& 0.01 &  0.1 &  1 &  2 &  5 \\
\hline
$x^*$  & 0.384 & 0.378 & 0.352 & 0.254 & 0.197 & 0.118 \\
\hline
$x^L$  & 0.284 & 0.279 & 0.259 & 0.186 & 0.143 & 0.085 \\
\hline
$g(\alpha)$ & 35.2\% &  35.5\% & 35.9\% & 36.6\% & 37.8\% & 38.8\%\\
\hline
\end{tabular}
\end{center}
\label{t:gain_fixrho}
\end{table}

\begin{figure}[htp]
\begin{center}
\includegraphics[width=0.42\textwidth]{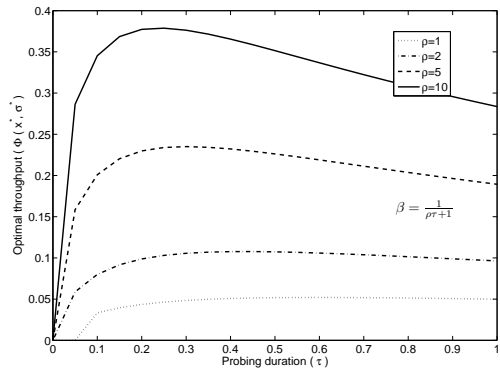}
\end{center}
\caption{Throughput vs. training time $\tau$} \label{fig:tau}
\end{figure}

We also examine the performance of distributed opportunistic
scheduling as a function of the training time $\tau$. According to linear
estimation theory, it has been shown that the error variance $\beta$ and the
training time $\tau$ have the following relationship~\cite{dong2007}:
\begin{equation}
\beta = \frac{1}{\rho \tau +1}. \label{eq:betau}
\end{equation}
Using \sr{eq:betau} in the simulation, we plot throughput
performance of the distributed opportunistic scheduling as a function of the
training time $\tau$ in Fig.~\ref{fig:tau}. It is clear  that there exists an optimal training time which balances the tradeoff between better estimation accuracy and loss of transmission time. It can
also be observed that when the average SNR $\rho$ increases, the
optimal training time decreases.

\section{Conclusion} \label{sec:con}
In this paper, we have generalized the study in~\cite{zgz:mobihoc} to investigate distributed opportunistic scheduling under noisy channel estimation. For such cases, we have proposed that the transmission rate backs off on the estimated rate so as to reduce the outage probability. We have showed that the optimal scheduling policy has a threshold structure, but the threshold turns out to be a function of the variance of the estimation error, and is a functional of the backoff rate function. Since the optimal backoff is analytically intractable, we have proposed a suboptimal linear scheme that backs off on the estimated SNR (and hence the rate). The corresponding optimal backoff ratios and rate thresholds can be obtained using an iterative algorithm based on fractional maximization. Simulation results indicate that DOS still achieves significant throughput gain in the presence of noisy channel estimation, especially in the low SNR region. In addition, we have observed that the performance loss of DOS due to the imperfect channel estimation is less than that of the schemes that do not use channel-aware scheduling, indicating that the devised DOS is more robust against noisy channel estimation.

\bibliographystyle{IEEEtranS}
\bibliography{reference}
\end{document}